# Transport inefficiency in branched-out mesoscopic networks: An analog of the Braess paradox

M. G. Pala[1,∗], S. Baltazar[1], P. Liu[2], H. Sellier[2], B. Hackens[3], F. Martins[3], V. Bayot[2,3], X. Wallart[4], L. Desplanque[4], and S. Huant[2,∗]

1: IMEP-LAHC, Grenoble INP, Minatec, BP 257, 38016 Grenoble, France
2: Institut Néel, CNRS and Université Joseph Fourier, BP 166, 38042 Grenoble, France
3: CERMIN, DICE Lab, UCL, B-1348 Louvain-la-Neuve, Belgium
4: IEMN, UMR CNRS 8520, UST Lille, BP 60069, 59652 Villeneuve d'Ascq, France

We present evidence for a counter-intuitive behavior of semiconductor mesoscopic networks that is the analog of the Braess paradox encountered in classical networks. A numerical simulation of quantum transport in a two-branch mesoscopic network reveals that adding a third branch can paradoxically induce transport inefficiency that manifests itself in a sizable conductance drop of the network. A scanning-probe experiment using a biased tip to modulate the transmission of one branch in the network reveals the occurrence of this paradox by mapping the conductance variation as a function of the tip voltage and position.

The Braess paradox occurring in congested traffic [1] or other classical networks [2,3] is a counter-intuitive behavior where the overall network performance can paradoxically degrade after addition of an extra capacity. For example, adding a new road to a complex network where travelers selfishly choose their own way, being just inspired by their knowledge of the global or local network structure, can lead to a longer transport time at equilibrium for all travelers. In turn, blocking certain streets in complex networks of congested cities can partially and counter-intuitively relieve congestion [4]. This original Braess paradox [1] can be explained within the frame of game theory since it may occur that the equilibrium point of a many-player game where each participant is able to anticipate the strategy of rivals, i.e., the Nash equilibrium [5], may counter-intuitively lead to a poor compromise for all players. Hence, the Braess paradox can be viewed as a non-cooperative game for which Nash equilibrium does not follow a simple "80-20" rule, i.e., is not "Pareto efficient" [6]. Its macroscopic analogs in classical physics, such as e.g. the electrical [2] or mechanical [3] Braess paradoxes, are then just seen as extensions of the original road-network paradox.

     Apart from a hint that a related paradox contributes to the magnetism in nanostructured ferromagnets [7], there is to our knowledge no report on a possible analog of the Braess paradox in quantum physics where interactions are mostly driven by phase coherence and non-locality effects, and elementary excitations obey specific non-poissonian statistics. In this letter, we focus our attention on the occurrence of a quantum analog of the Braess paradox and discover, both numerically and experimentally, that such an anolog can indeed occur in semiconductor mesoscopic, i.e. phase-coherent, networks.

     For our demonstration, we concentrate on mesoscopic samples with a spatial extension of hundreds of nanometers, such that many conducting channels contribute to the total current, but the device lengths remain smaller than the coherence length, which is of several microns for high-mobility semiconductor structures at low temperature [8]. A reason for focusing on large mesoscopic samples at low temperature is that the large number of transverse modes involved in transport [9,10] allows to distinguish the occurrence of an analog of the Braess paradox from merely phase-coherent or size effects which are important only when the structures are weakly coupled to the leads or a few transverse modes propagate through them [11]. In addition, by considering a large

∗: E-mail to pala@minatec.inpg.fr or serge.huant@grenoble.cnrs.fr

number of modes $N$ the broadening of each energy level is about $N$ times the level spacing and the density of states becomes almost featureless [12].

To explore the possibility of a mesoscopic Braess paradox, we propose a simple two-path network depicted in Figs. 1(a-c). The basis network takes the form of a 1.0 μm x 1.6 μm rectangular corral connected to left (source) and right (drain) ohmic leads via two planar wires (openings) of width $W_0$ = 300 nm (Fig. 1(a)). The upper and lower wires of width $W_1 = W_2$ = 100 nm are chosen narrower than the openings so as to behave as constrictions for propagating electrons. A central wire directly connecting source to drain and opening an alternative third path to bypass the central "antidot" (Figs. 1(b) and 1(c)) should intuitively result in an increased total current. We show below that it is exactly the opposite behavior that actually takes place. We address the influence of the third route by progressively increasing its width $W_3$ beyond that of the two openings. Note that the choice of narrow constrictions ensures that the electron flow is congested. Indeed, in a system where electrons can be back-scattered solely by the walls defining the structure geometry, a sufficient condition to reach congestion is obtained when the number of conducting modes allowed by internal constrictions is smaller than the number of conducting modes in the external openings, which implies $W_1+W_2<W_0$.

We first evaluate the transport properties of such a system by means of an exact numerical approach based on the Keldysh Green's function formalism [13]. For typical InGaAs/InAlAs heterostructures [14,15] (such as those considered later in this letter) the electron concentration in the two-dimensional electron gas (2DEG) results in a Fermi energy $E_F$ of several tens of meV, which, assuming an effective mass of 0.04 $m_0$, corresponds to a Fermi wavelength $\lambda_F$ of tens of nm. This implies the injection/detection of tens of conducting modes through the openings. We consider the device in the linear transport regime and perform a thermal average around $E_F$ at the temperature $T = 4.2K$. A square mesh with 2.5 nm long sides is used in our calculations. The Green's function of the system is described in the real-space representation that allows us to include all possible conducting and evanescent modes. Moreover, in order to reduce the computational time and memory requirement we adopt a recursive strategy based on the Dyson equation [16-18].

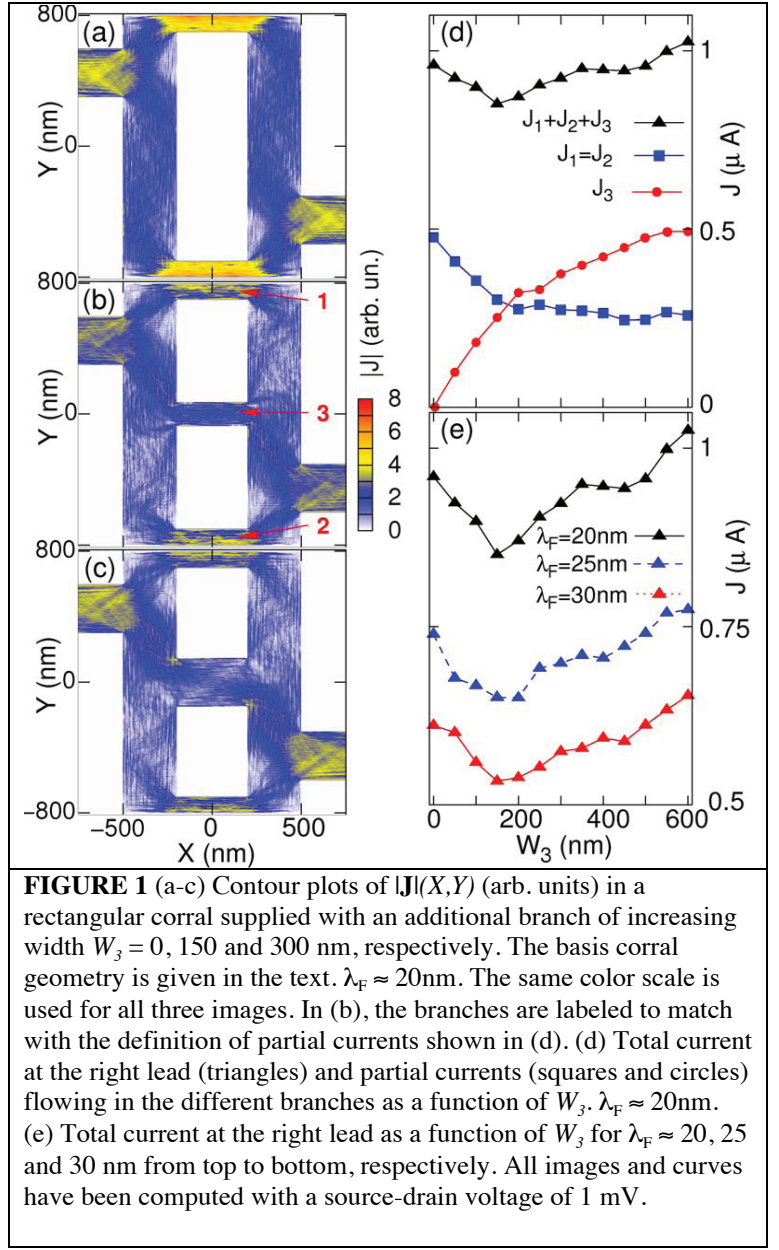

**FIGURE 1** (a-c) Contour plots of |**J**|(X,Y) (arb. units) in a rectangular corral supplied with an additional branch of increasing width $W_3$ = 0, 150 and 300 nm, respectively. The basis corral geometry is given in the text. $\lambda_F \approx$ 20nm. The same color scale is used for all three images. In (b), the branches are labeled to match with the definition of partial currents shown in (d). (d) Total current at the right lead (triangles) and partial currents (squares and circles) flowing in the different branches as a function of $W_3$. $\lambda_F \approx$ 20nm. (e) Total current at the right lead as a function of $W_3$ for $\lambda_F \approx$ 20, 25 and 30 nm from top to bottom, respectively. All images and curves have been computed with a source-drain voltage of 1 mV.

Fig. 1(d) demonstrates the occurrence of a Braess-like paradoxical behavior by showing (triangles) the current across the structure $J(W_3)$ as a function of the third channel width. Even if intuition suggests that enlarging the central wire should increase the total current, we find that over a large width range, $J(W_3)$ monotonically decreases until loosing more than 10 percents at around $W_3$=150 nm. This correction is comparable to the Braess paradox in classical systems [2,3]. Furthermore, for $W_3$ larger than the opening width, $J(W_3)$ slowly increases and eventually overcomes $J(0)$ when $W_3$ is large enough ($W_3$>500nm) to strongly reduce electron reflections due to the antidot. This counter-intuitive behavior can be attributed neither to resonant effects, because of the large number of modes participating in transport, nor to quantum conductance fluctuations, which we compute to be 1 or 2% of the average conductance.

More physical insight into the $W_3$ dependence of the net current is obtained by mapping the spatial distribution of the current density $|\mathbf{J}|(X,Y)$ for increasing $W_3$. Congested electron flows are clearly observed for the basis network in Fig. 1(a) where current maxima are located inside the upper and lower wires. Now, adding a third path to electrons via the central wire modifies the current spatial distribution and hence reduces the current flowing through the lateral wires by providing carriers an alternative path to propagate from source to drain. However, as long as the central wire is not large enough to permit a significant direct coupling between left and right contacts, the net current at the right lead decreases as a function of $W_3$. In Fig. 1(b) the central wire is strongly coupled both to the upper and lower constrictions, showing that when the paradox occurs electrons can experience closed orbits inside the corral. Fig. 1(c) shows the cases where the coupling between the central wire and the two openings increases and becomes dominant with respect to the competing coupling between internal paths.

Such a qualitative picture is clarified in Fig. 1(d) where we calculate the currents $J_1$, $J_2$, and $J_3$ flowing in the middle of the device ($X = 0$) through the upper, lower, and central branches, respectively. The initial effect of opening the central wire is to drastically reduce $J_1$ and $J_2$, whereas $J_3$ is initially too small to compensate for the loss in drain current. As soon as $W_3$ increases and $J_3$ becomes comparable to $J_1 = J_2$ for $W_3$ around 150 nm, the total current stops decreasing since the direct coupling between the left and right openings becomes dominant and finally the paradox is lifted. The fact that the minimum current condition occurs when the partial currents in the three branches are identical suggests a direct analogy with the classical Braess paradox for which the maximum penalty occurs when the maximal symmetry permits a minimum of independent parameters [3].

A demonstration of the robustness of the effect is shown in Fig. 1(e), where we consider three different Fermi energies corresponding to $\lambda_F \approx 20$, 25 and 30 nm. The current exhibits a similar behavior as a function of $W_3$, showing a clear reduction for a large range of the third wire width, independently of the Fermi energy. This confirms that the current decrease is not an interference effect.

We now turn to an experimental demonstration of the paradox introduced above. Due to the difficulty in fabricating several devices differing only in the central wire width we propose an approach based on scanning-gate microscopy (SGM). SGM uses the biased tip of a cryogenic atomic-force microscope (AFM) to scan over a buried semiconductor device and alter locally its electrostatic potential and hence its conductance [19,20]. SGM has been previously used to image and manipulate electron transport in various nanostructures [14,15,17-26]; for a review, see [27]. Here, we use SGM to tune at will the electron transmission through the additional branch in a branched-out network.

The device is shown in Fig. 2(a). It has been patterned from an InGaAs/InAlAs heterostructure forming a 2DEG 42nm below the surface. The carrier density is

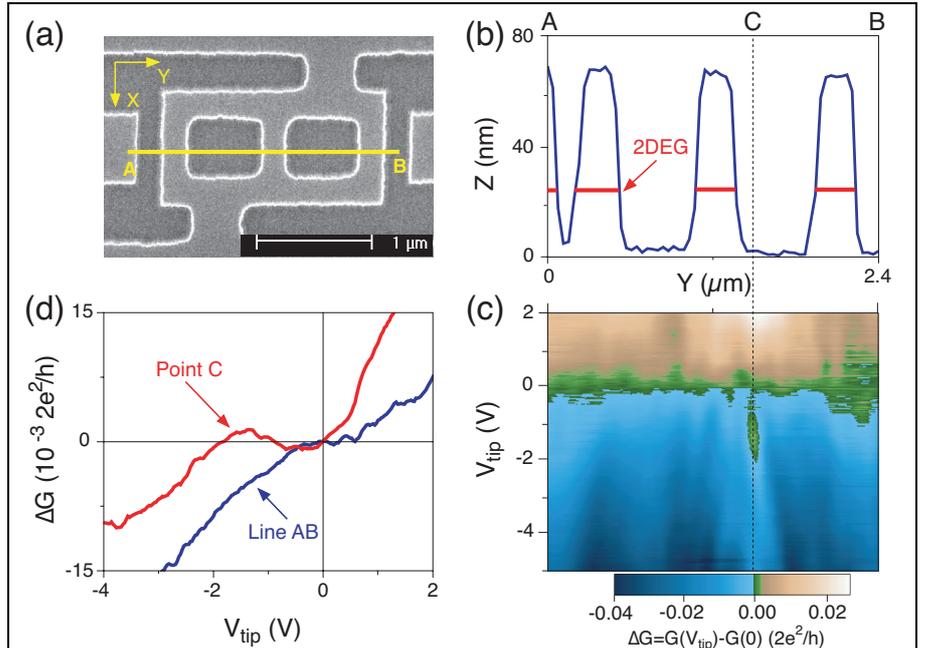

**FIGURE 2** (a) Scanning electron micrograph of the branched-out mesoscopic network. (b) AFM topographic profile along the AB line. (c) Conductance changes as a function of $V_{tip}$ recorded with the tip scanning above the AB line. (d) Conductance change extracted from (c) at point C compared with the average change over the entire AB line.

$3.5\times10^{11}$cm$^{-2}$ ($\lambda_F$=42 nm) and the mobility is 100000cm$^2$V$^{-1}$s$^{-1}$ at 4.2K. The device dimensions are close to those used in the simulations: the overall size is 1.0 x 1.7 μm$^2$, all arms are around 220nm wide and the leads to the reservoirs are 420nm wide. The sample is mounted in a cryogenic AFM thermalized at 4.2K in helium gas. The topography is obtained with a conductive AFM tip attached to a quartz tuning fork used as a force sensor [25,27]. Fig. 2(b) shows the sample profile measured along the AB line.

Immediately after recording the topographic profile, the tip is lifted to 100nm above the surface and scanned at constant height along the same AB line while the conductance $G(Y_{tip})$ is recorded as function of the tip position $Y_{tip}$ at fixed tip voltage $V_{tip}$. The conductance profiles obtained for different $V_{tip}$ are gathered on a color plot in Fig. 2(c). Here, the $V_{tip}=0$ profile (related to a non-perfect tip [27,28]) has been subtracted in order to focus on the conductance changes $\Delta G(Y_{tip},V_{tip}) = G(Y_{tip},V_{tip})-G(Y_{tip},0)$ induced by $V_{tip}$ (the conductance without tip is around $2e^2/h$). Fig. 2(c) shows that the conductance for each $Y_{tip}$ along AB decreases monotonically for negative $V_{tip}$ and increases monotonically for positive, with a distinctive exception at point C, located in the vicinity of the central arm, where the conductance shows a clear non-monotonic behavior near $V_{tip} = -1V$. This region has been highlighted by changing abruptly the color scale at $\Delta G =0$. The voltage sweep at this particular location is plotted in Fig. 2(d) and compared with the average of all sweeps recorded along the AB line. The decreasing conductance with increasing $V_{tip}$ observed at point C in the range [-1.3V, -0.3V] with an amplitude around $2.5 \times 10^{-3}$ $2e^2/h$ is much larger than the small conductance fluctuations observed at other locations. Since decreasing $V_{tip}$ corresponds to decreasing the wire transmission, this counter-intuitive increase in current is taken as evidence for a Braess-like paradox.

Fig. 2(c) highlights the particular role played by the central branch in the network since the paradoxical conductance increase is observed only when the negatively biased tip scans around point C located close to the central branch, not when it scans over a lateral branch. Repeated measurements after additional thermal cycles showed the robustness of the observed phenomenon. However, we could notice that the exact location around the central channel, $V_{tip}$ range and amplitude of the conduction increase shown in Fig. 2(c) all changed with the sample history. This is typical for semiconductor mesoscopic devices and is a consequence of the fluctuating potential landscape probed by the 2DEG due, e.g., to residual charge traps that can distort SGM images [27]. In addition, the SGM tip can progressively be eroded or polluted after long-term measurements so that the electrostatic and topographic tip apex may no longer match [27]. These are likely the reasons why point C in Figs. 2(b-c) is shifted by 270 nm to the right hand side of the central path.

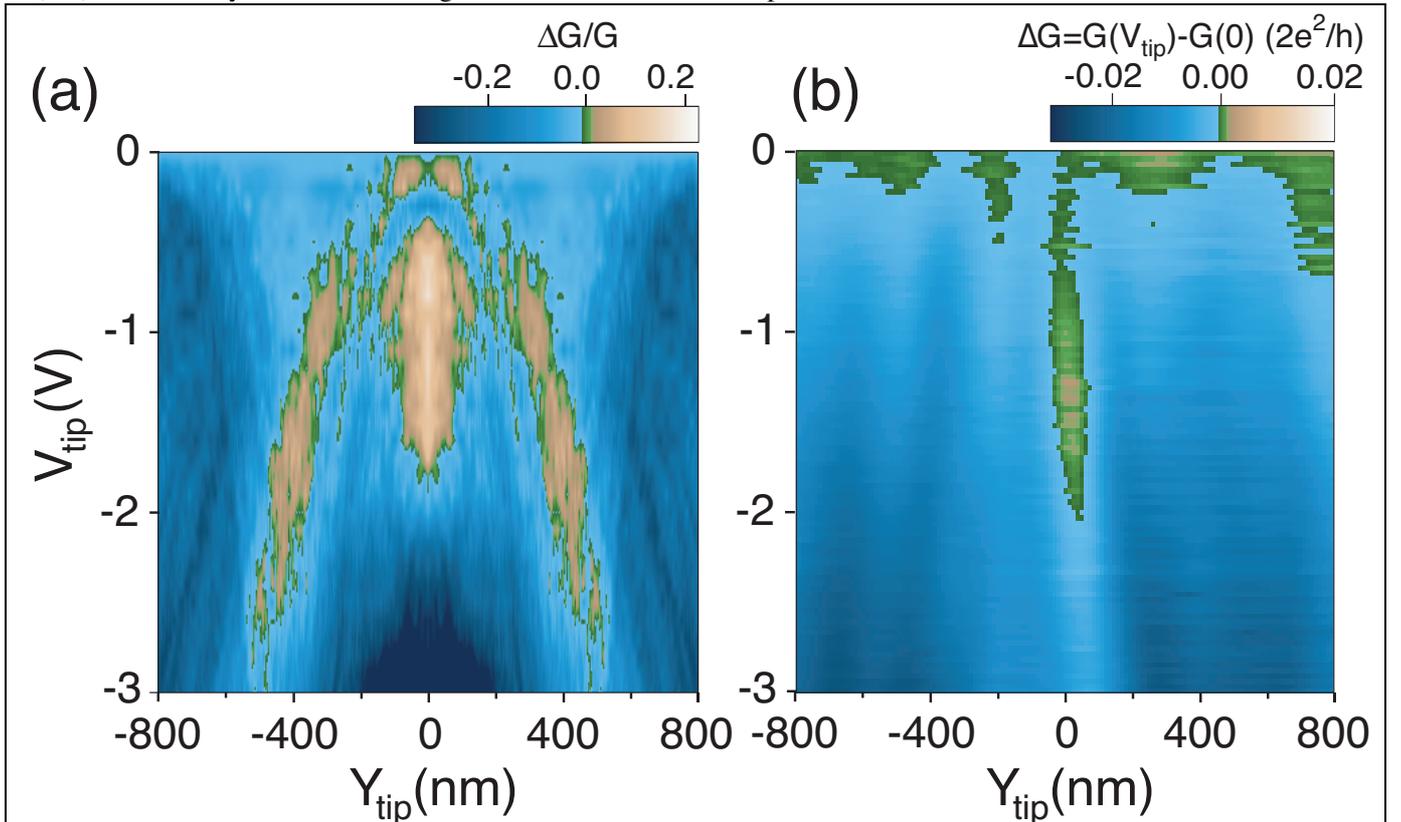

**FIGURE 3** (a) Conductance variation map as a function of $V_{tip}$ and $Y_{tip}$ in the middle of a Braess-paradox device ($X_{tip} = 0$). The corral length and width are 0.8 μm and 1.6 μm. The width of the openings, lateral arms, and central branch are 340 nm, 140 nm, and 160 nm, respectively. $V_{DS}= 1$ mV. $\lambda_F =47$ nm. (b) A zoom on the experimental image of Fig. 2(c) over the same voltage-position window as in (a). The data have been centered on point C to account for the misfit of 270 nm between topographic and electrostatic images discussed in the text.

The experiment in Fig. 2(c) can be simulated by computing the device conductance [17,18]. This is done in Fig. 3(a), where the tip-induced conductance variation $\Delta G/G$ is mapped as a function of $Y_{tip}$ and $V_{tip}$ at fixed $X_{tip}$ = 0. Structure dimensions are chosen to fit Fig. 2(a) [29]: see Fig. 4. The tip potential is simulated by a point-like gate voltage located 100 nm above the 2DEG, which results in a spatial extension of the perturbing potential of about 400 nm for $V_{tip}$= -1 V. A clear conductance increase occurs around $V_{tip}$= -1 V when the electrostatic potential due to the scanning gate depletes the central wire and allows electrons to flow through the two lateral wires only.

Clearly, simulations reproduce qualitatively the experimental result shown in Fig. 2(c) that is zoomed on in Fig. 3(b) over the same $(Y_{tip},V_{tip})$ window. The lateral stripes of positive conductance corrections in Fig. 3(a) are due to the non-vanishing tail of the tip-induced potential that for certain values of $Y_{tip}$ and $V_{tip}$ can improve the direct coupling between source and drain via the central wire.

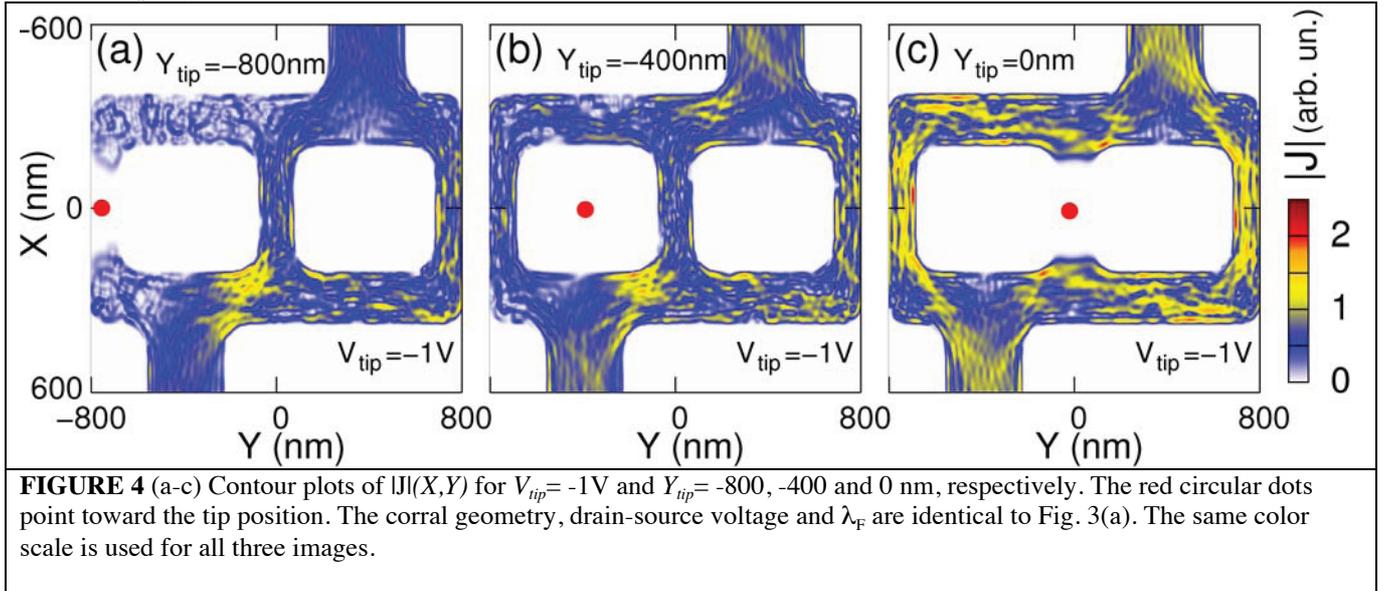

**FIGURE 4** (a-c) Contour plots of $|J|(X,Y)$ for $V_{tip}$= -1V and $Y_{tip}$= -800, -400 and 0 nm, respectively. The red circular dots point toward the tip position. The corral geometry, drain-source voltage and $\lambda_F$ are identical to Fig. 3(a). The same color scale is used for all three images.

Figs. 4(a-c) help to interpret this finding by showing how the current spatially redistributes for different tip positions. In Fig. 4(a) the SGM tip is located over the lateral wire and, closing the passage of electrons through the first wire, induces enhanced internal reflections (visible in the left region) that result in a negative conductance variation, whereas, in Fig. 4(b) the tip is over the antidot and induces a negligible $\Delta G/G$. The counter-intuitive behavior of the Braess paradox is recovered when comparing Figs. 4(b) and (c). When the tip is located over the central wire such that only two macroscopic paths are allowed to carriers (case (c)), a larger current flows through the device with respect to both the (a) case, where a lateral channel is closed, and the (b) case, where three paths are allowed with possible internal closed orbits. This confirms the particular role played by the central channel in the occurrence of the mesoscopic Braess paradox that was already mentioned in the experimental part.

In conclusion, we have discovered a mesoscopic analog of the Braess paradox, which was known so far only for macroscopic networks governed by classical physics. Our findings raise fundamental issues. For instance, can our microscopic explanation in terms of current redistribution within the network be reconciled with game theory widely invoked to explain the classical paradox? Can the paradox manifest itself in other coherent systems such as plasmonic interferometers [30]? We hope the present letter will attract interest towards such studies.

SH acknowledges a fruitful discussion on the Braess paradox with B. Cleuren, Hasselt University that has stimulated the present study. We thank G. Bachelier, H. Courtois, A. Cresti, M. Governale, A. Drezet and B. Saépé for helpful discussions. B.H. is research associate and F.M. is a postdoctoral researcher with the Belgian FRS-FNRS. This work has been supported by FRFC grant no. 2.4.546.08.F and FNRS grant no. 1.5.044.07.F, by the Belgian Science Policy (Program IAP-6 / 42) and by the French Agence Nationale de la Recherche ("MICATEC" project). PL and VB acknowledge support from the Grenoble Nanosciences Foundation ("Scanning-Gate Nanoelectronics" project).